\documentclass[sigconf, nonacm]{acmart}
\AtBeginDocument{%
  }

\usepackage[utf8]{inputenc}
\usepackage{tcolorbox}
\usepackage{tikz}
\usetikzlibrary{positioning, arrows.meta}
\usepackage{caption}
\usepackage{courier}
\usepackage{setspace}
\usepackage{booktabs}
\usepackage{graphicx}
\usepackage{multirow}

\newcommand{\negcell}[1]{\textcolor{red}{#1}}
\newcommand{\sig}{\textsuperscript{*}}

\setcopyright{acmlicensed}
\copyrightyear{2025}
\acmYear{2025}
\setcopyright{acmlicensed}\acmConference[SIGIR-AP 2025]{Proceedings of the 2025 Annual International ACM SIGIR Conference on Research and Development in Information Retrieval in the Asia Pacific Region}{December 7--10, 2025}{Xi'an, China}
\acmBooktitle{Proceedings of the 2025 Annual International ACM SIGIR Conference on Research and Development in Information Retrieval in the Asia Pacific Region (SIGIR-AP 2025), December 7--10, 2025, Xi'an, China}
\acmDOI{10.1145/3767695.3769493}
\acmISBN{979-8-4007-2218-9/2025/12}

\begin{document}

\title{Do Large Language Models Favor Recent Content? \\ A Study on Recency Bias in LLM-Based Reranking}


\author{Hanpei Fang}
\affiliation{%
  \institution{Waseda University}
  \city{Tokyo}
  \country{Japan}}
\email{hanpeifang@ruri.waseda.jp}

\author{Sijie Tao}
\affiliation{%
  \institution{Waseda University}
  \city{Tokyo}
  \country{Japan}
}
\email{tsjmailbox@ruri.waseda.jp}

\author{Nuo Chen}
\affiliation{%
 \institution{The Hong Kong Polytechnic University}
 \city{Hong Kong}
 \country{China}
}
\email{pleviumtan@outlook.com}

\author{Kai-Xin Chang}
\affiliation{%
  \institution{Waseda University}
  \city{Tokyo}
  \country{Japan}}
\email{victorchang@toki.waseda.jp}

\author{Tetsuya Sakai}
\affiliation{%
  \institution{Waseda University}
  \city{Tokyo}
  \country{Japan}}
\email{tetsuyasakai@acm.org}

\renewcommand{\shortauthors}{Hanpei Fang, Sijie Tao, Nuo Chen, Kai-Xin Chang, and Tetsuya Sakai}

\begin{abstract}

Large language models (LLMs) are increasingly deployed in information systems, including being used as second-stage rerankers in information retrieval pipelines, yet their susceptibility to recency bias has received little attention. We investigate whether LLMs implicitly favour newer documents by prepending artificial publication dates to passages in the TREC Deep Learning passage retrieval collections in 2021 (DL21) and 2022 (DL22). Across seven models, GPT-3.5-turbo, GPT-4o, GPT-4, LLaMA-3 8B/70B, and Qwen-2.5 7B/72B, ``fresh'' passages are consistently promoted, shifting the Top-10's mean publication year forward by up to 4.78 years and moving individual items by as many as 95 ranks in our listwise reranking experiments. Although larger models attenuate the effect, none eliminate it. We also observe that the preference of LLMs between two passages with an identical relevance level can be reversed by up to 25\% on average after date injection in our pairwise preference experiments. These findings provide quantitative evidence of a pervasive \textbf{recency bias} in LLMs and highlight the importance of effective bias-mitigation strategies.

\end{abstract}




\maketitle

\section{Introduction}
Large language models (LLMs) have been adopted throughout the information retrieval (IR) pipeline in an ever more diverse range of roles \cite{10.1145/3642979.3643006, zhu2024largelanguagemodelsinformation, li2024llms}. Beyond the ``classic'' tasks of query expansion and rewriting \cite{gao-etal-2023-precise, wang-etal-2023-query2doc}, recent studies have explored LLMs for retrieval augmented generation (RAG) \cite{gao2023retrieval, jiang-etal-2023-active, 10.1145/3726302.3730078}, large-scale relevance annotation \cite{upadhyay2024umbrela, upadhyay2024large, upadhyay2024llms, takehi2025llm}, and second-stage reranking of search-engine result pages (SERPs) produced by sparse or dense first-stage retrievers \cite{sun-etal-2023-chatgpt, gangi-reddy-etal-2024-first, ma2023zero, zhuang2024setwise, qin2024large, 10.1145/3696410.3714658, pradeep2023rankvicunazeroshotlistwisedocument, pradeep2023rankzephyr, sharifymoghaddam2025rankllmpythonpackagereranking}. While these applications show promise, recent work has surfaced several intrinsic weaknesses.

One primary concern is that LLMs may inherit and even amplify social \cite{abid2021persistent, bang-etal-2024-measuring, omiye2023large, zhao2024comparative, soundararajan-delany-2024-investigating, gallegos-etal-2024-bias, hu2024} and cognitive \cite{chen2024ai, eicher2024reducing, enke2023cognitive, itzhak2024instructed, schmidgall2024addressing, 10.1145/3708884} biases embedded in their pre-training corpora, thereby propagating undesirable effects into downstream IR tasks. A second issue is prompt sensitivity. Seemingly innocuous prompt tweaks can produce systematic errors. Arabzadeh~et al. \cite{arabzadeh2025human} show that even minor prompt changes sharply skew graded relevance judgments, whereas Alaofi~et al. \cite{alaofi2024llms} demonstrate that simply copying the query text into the document, a long-standing search engine optimization (SEO) strategy, causes widely used LLMs to overestimate relevance. Such vulnerabilities raise doubts about the reliability of LLM-driven components in end-to-end IR systems.

Another long-standing SEO strategy is to exploit \emph{recency signals}. Search engines often reward pages that appear freshly updated, assuming newer content better satisfies users' needs \cite{10.1145/1718487.1718490, 10.1145/1718487.1718489, campos2014survey}. However, the majority of ``updates'' are triggered by minor modifications \cite{10.1145/1498759.1498837, 10.1002/spe.577}, while leaving the substantive text untouched, which nevertheless reset the ``Last updated on'' or ``Published on'' timestamp. While these \emph{pseudo-fresh} pages can still climb traditional rankings, it remains unclear whether rankers powered by LLMs exhibit the same vulnerability.

In this paper, we investigate whether a \textbf{recency bias} exists in LLM-based IR systems and, if so, how strongly it distorts ranking outcomes. Our study targets passage reranking and poses a single guiding question: \emph{Do LLMs systematically prefer newer content when acting as search rerankers?}

To answer this, we devise a listwise reranking experiment that injects artificial publication dates into candidate passages. We observe how seven LLMs from three different providers adjust their rankings, covering both lightweight, cost-effective models and heavy, high-capacity alternatives, and we introduce multiple evaluation metrics that quantify any resulting temporal shifts. To gauge the strength of these recency cues more precisely, we complement the listwise study with pairwise preference tests on four open-source models. The key contributions of this work are:
\begin{description}
    \item[Recency Bias Revealed.] We show that recency bias is pervasive across LLM-based rerankers: every model we test systematically promotes passages that merely appear ``fresh''. Larger models alleviate, but never eliminate, this effect.
    \item[Diagnostic Framework.] We introduce a reranking-based testing methodology together with a complementary metric suite that quantifies how strongly temporal cues sway LLM decisions.
    \item[Broader Implications.] By exposing recency bias, we invite the community to probe and ultimately mitigate the wider spectrum of hidden intent biases lurking in LLM-centric IR systems.
\end{description}

\section{Related Works}

\subsection{LLM-Based Reranking}

Recent advances have woven LLMs into almost every layer of the IR stack, including second-stage text reranking and relevance assessment. In second-stage reranking, an LLM receives the top-k results retrieved by a sparse or dense first-stage retriever and returns a refined ordering of the SERP \cite{sun-etal-2023-chatgpt, zhuang2024setwise, qin2024large}. The prevailing method is listwise reranking \cite{tang-etal-2024-found, gangi-reddy-etal-2024-first, ma2023zero, 10.1145/3696410.3714658}: a chunk of documents is fed to the model in a single prompt, and the model outputs a reordered list. Dedicated systems such as RankGPT \cite{sun-etal-2023-chatgpt}, RankVicuna \cite{pradeep2023rankvicunazeroshotlistwisedocument}, RankZephyr \cite{pradeep2023rankzephyr} and RankLLM \cite{sharifymoghaddam2025rankllmpythonpackagereranking} further push performance.

Because listwise prompts quickly exhaust the model's context window, researchers typically adopt a sliding-window strategy \cite{sun-etal-2023-chatgpt}: overlapping segments of the initial ranking are processed in turn so that even low-ranked (tail) documents are scored and can be promoted if relevant. Building on this paradigm, we design a listwise reranking experiment that injects synthetic timestamps to probe whether and how strongly LLMs exhibit recency bias.

Pairwise reranking \cite{qin2024large} can be viewed as a special case of listwise reranking with a window size of two. Leveraging this connection, we supplement our listwise study with pairwise preference tests on four open-source models, yielding finer-grained evidence of how temporal cues sway LLMs' preferences.

\subsection{LLM-Based Relevance Assessment}
Constructing IR test collections hinges on relevance assessment, a painstaking, expensive bottleneck. Inspired by LLM successes in other domains, recent studies have explored replacing or supplementing human judges with LLMs \cite{upadhyay2024umbrela, upadhyay2024large, upadhyay2024llms, abbasiantaeb2024uselargelanguagemodels, 10.1145/3578337.3605136, 10.1145/3539618.3592032, 10.1145/3626772.3657707}, and others have sought to keep humans in the loop while shrinking their workload \cite{takehi2025llm}. LLM-based assessors offer two clear advantages: (i) each document can be labelled independently, eliminating inter-document leakage; and (ii) annotation cost and turnaround time drop by orders of magnitude compared with human assessors. Nonetheless, recent evidence highlights serious reliability pitfalls when LLMs serve as automatic judges \cite{clarke2024llm}.

\subsection{Vulnerability in Large Language Models}
LLMs possess intrinsic weaknesses that can undermine downstream applications, such as text reranking and relevance assessment. Hallucination \cite{huang2025survey} is the most widely discussed flaw, where LLMs generate seemingly factual yet unfounded content. Beyond hallucination, Wallat~et al. \cite{wallat2024correctnessfaithfulnessragattributions} emphasise the need for faithfulness in RAG tasks: answers must be grounded strictly in retrieved evidence rather than the model conjectures. They probe this weakness by injecting adversarial statements derived from an LLM's initial answer into documents, whether random, relevant but uncited, or previously cited for other reasons, and then regenerating the answer to see if the fabricated content is incorporated.

Bias is another crucial concern when using LLMs in downstream tasks \cite{10.1145/3637528.3671458, wang2025biasamplificationragpoisoning}. Fine-grained audits uncover pervasive social biases \cite{gallegos-etal-2024-bias}, including gender \cite{zhao2024comparative, soundararajan-delany-2024-investigating}, racial \cite{omiye2023large}, political \cite{bang-etal-2024-measuring} and religious \cite{abid2021persistent} stereotypes, in LLMs' text generation and ranking outputs, as well as cognitive biases \cite{eicher2024reducing, chen2024ai, itzhak2024instructed, schmidgall2024addressing}. Chen~et al. \cite{chen2024ai}, for instance, demonstrate that the threshold priming effect skews LLM relevance judgments.

Besides these intrinsic issues, LLMs are also susceptible to simple adversarial manipulations. Alaofi~et al. \cite{alaofi2024llms} demonstrate that embedding the raw query into an otherwise irrelevant passage often suffices to elicit a ``highly relevant'' label in relevance judgments, exposing a basic keyword-stuffing vulnerability. Arabzadeh~et al. \cite{arabzadeh2025human} systematically analysed the sensitivity in LLM-based relevance judgement, where they show that minor prompt variations alone can swing graded relevance judgments, while irrelevant or distracting context can also erode performance \cite{wu2024easilyirrelevantinputsskew, 10.5555/3618408.3619699, yang2025llmreasoningdistractedirrelevant}. 

Motivated by these findings and inspired by the adversarial setups of Wallat~et al.\cite{wallat2024correctnessfaithfulnessragattributions} and Alaofi~et al.\cite{alaofi2024llms}, this paper turns to an under-explored weakness, \textbf{recency bias}. We isolate timestamps as the sole manipulated variable, injecting artificial publication dates to measure how strongly temporal signals distort LLM-based reranking.

\subsection{Search Engine Optimization Strategy}
Search engine optimization (SEO) strategy refers to any deliberate tactic that boosts a page's ranking. A classic black-hat technique, keyword stuffing, has already been shown to deceive LLM-based relevance assessors \cite{alaofi2024llms}. Modern search engines, however, also reward freshness: recency-aware ranking models weigh timestamps, update frequency, and other temporal signals to satisfy time-sensitive queries \cite{10.1145/1718487.1718490,10.1145/1718487.1718489,campos2014survey}. Crucially, the majority of ``updates'' are triggered by minor modifications \cite{10.1145/1498759.1498837, 10.1002/spe.577}, including purely cosmetic edits such as fixing a typo, tweaking formatting, or making other negligible changes, while leaving the substantive content untouched; nevertheless, doing so rewrites the ``Last updated on'' or ``Published on'' timestamp. In this work, we ask whether LLM-based rerankers fall for the same ploy and quantify the magnitude of the resulting recency bias.

\section{Experiments}

\subsection{Test Collections, LLMs and Prompt}
Our listwise experiments use the passage retrieval test collections from the TREC 2021 Deep Learning Track (DL21) \cite{craswell2022overview} and the TREC 2022 Deep Learning Track (DL22) \cite{craswell2023overview}. We retain only queries with NIST human relevance judgments, yielding 53 DL21 queries and 76 DL22 queries. We evaluate seven LLMs, spanning three providers and two parameter scales:

\begin{description}
  \item[OpenAI:] GPT-3.5-turbo (1106), GPT-4 (0613) \cite{achiam2023gpt}, and GPT-4o (2024-05-13) \cite{hurst2024gpt}.
  \item[Meta AI:] LLaMA3-instruct-8B and LLaMA3-instruct-70B \cite{dubey2024llama}.
  \item[Alibaba Cloud:] Qwen2.5-7B and Qwen2.5-72B \cite{qwen2025qwen25technicalreport}.
\end{description}

All models are queried with identical decoding settings: \texttt{top\_p = 1.0}, \texttt{temperature = 0}, \texttt{frequency\_penalty = 0}, and \texttt{presence\_penalty = 0}. We adopt the RankZephyr \cite{pradeep2023rankzephyr} prompt for all models; the complete template appears in Figure~\ref{fig:ranking-prompt}.

\begin{figure}
\centering
\begin{tcolorbox}[colback=gray!5, colframe=black!50, boxrule=0.5pt,
arc=2pt, left=5pt, right=5pt, top=5pt, bottom=5pt]

\small\ttfamily
You are RankLLM, an intelligent assistant that can rank passages based on their relevancy to the query.

\

I will provide you with \{n\} passages, each indicated by a numerical identifier []. \\
Rank the passages based on their relevance to the search query: \{query\}.

[1] \{passage1\} 
\\\
[2] \{passage2\} 
\\\
...

Search Query: \{query\}

Rank the \{n\} passages above based on their relevance to the search query. \\
All the passages should be included and listed using identifiers, in descending order of relevance.

The output format should be [] > [], e.g., [4] > [2]. \\
Only respond with the ranking results, do not say any word or explain.

\end{tcolorbox}

\caption{The prompt used in our listwise reranking experiments for ranking passages based on relevance to a query, copied from RankZephyr \cite{pradeep2023rankzephyr}.}
\label{fig:ranking-prompt}
\end{figure}

For the pairwise preference experiments, due to cost reasons, we exclude proprietary models and restrict our analysis to DL21 passages with NIST human judgments, using a dedicated prompt shown in Figure~\ref{fig:preference-prompt}. Even within this constraint, the selected open-source models span approximately one order of magnitude in parameter count and originate from two independent providers, yielding a broad test bed that demonstrates recency bias is not confined to any single architecture.

\subsection{Listwise Reranking Experiment}
To probe recency effects, we rerank the BM25 \cite{robertson2009probabilistic} baseline with each LLM, once on the original passages and once after injecting artificial publication dates. As in most LLM-based listwise rerankers, we apply a sliding window strategy with a window size of 10 to respect the context limit.

Figure~\ref{fig:date-injection} illustrates the date injection procedure. For every query, we first rerank the top-100 passages from the BM25 baseline in their original form. We then prefix every passage in the resulting SERP with \texttt{``Published on: \{Date\}.''} where \texttt{\{Date\}} is in the format of \texttt{``YYYY/MM/DD''}. Specifically, to create a controlled test environment, the passage at Rank 100 receives the most recent timestamp, \texttt{``Published on 2025/01/01.''}, and each higher-ranked passage is dated exactly one year earlier, so the passage at Rank 1 receives \texttt{``1926/01/01''}. We rerun the identical reranking on the modified list. Comparing the two resulting SERPs reveals how strongly explicit temporal cues sway the ranking. If the date injection exerts an overwhelming influence, the SERP should be nearly reversed.

\begin{figure}[ht]
\centering
\begin{tikzpicture}[
  label/.style={anchor=east, minimum height=0.7cm, font=\ttfamily\footnotesize},
  lbox/.style={rectangle, draw=black, minimum width=0.8cm, minimum height=0.7cm, font=\ttfamily\scriptsize, align=left},
  rbox/.style={rectangle, draw=black, minimum width=3.9cm, minimum height=0.7cm, font=\ttfamily\scriptsize, align=left},
  arrow/.style={-Stealth, thick, shorten >=2pt, shorten <=2pt},
  node distance=0.2cm
]

\node[label] (label_1) {\textbf{Rank 1:}};
\node[lbox, right=of label_1] (bef_1) {\textbf{P1}};
\node[rbox, right=1.9cm of bef_1] (inj_1) {\textbf{\textcolor{red}{Published on 1926/01/01.} P1}};

\node[label, below=of label_1] (label_2) {\textbf{...}};
\node[lbox, below=of bef_1] (bef_2) {\textbf{...}};
\node[rbox, below=of inj_1] (inj_2) {\textbf{...}};

\node[label, below=of label_2] (label_3) {\textbf{Rank 99:}};
\node[lbox, below=of bef_2] (bef_3) {\textbf{P99}};
\node[rbox, below=of inj_2] (inj_3) {\textbf{\textcolor{red}{Published on 2024/01/01.} P99}};

\node[label, below=of label_3] (label_4) {\textbf{Rank 100:}};
\node[lbox, below=of bef_3] (bef_4) {\textbf{P100}};
\node[rbox, below=of inj_3] (inj_4) {\textbf{\textcolor{red}{Published on 2025/01/01.} P100}};

\draw[arrow] (bef_1.east) -- node[midway, above=1pt,
                font=\footnotesize\bfseries]{Date injection} (inj_1.west);
\draw[arrow] (bef_3.east) -- (inj_3.west);
\draw[arrow] (bef_4.east) -- (inj_4.west);

\end{tikzpicture}

\caption{Date injection strategy: the passage at Rank 100 gets 2025/01/01, and each higher-ranked passage receives a date exactly one year earlier. P1-P100 denote the passage contents, and dates are injected as prefixes.}
\label{fig:date-injection}
\end{figure}

\subsection{Pairwise Preference Experiment}
To quantify the direct effect of temporal cues on passage preference, we conduct a pairwise test using the prompt shown in Figure~\ref{fig:preference-prompt}. For each query, we group the human-judged passages by relevance level (0, 1, 2, and 3) and generate all unordered pairs within each group.

In the first round, the LLM selects the more relevant passage in each pair, establishing a baseline preference. We then apply a controlled treatment (Figure~\ref{fig:date-injection-preference}): the initially preferred passage is prefixed with an antiquated date, \texttt{1980/01/01}, while the other passage receives a fresh date, \texttt{2025/01/01}. Dates are injected exactly as in the listwise experiment. Presenting the same prompt in the same order, we record the new preference. Because both passages are equally relevant by human judgment, any systematic preference reversal directly estimates the strength of recency bias.

\begin{figure}
\centering
\begin{tcolorbox}[colback=gray!5, colframe=black!50, boxrule=0.5pt,
arc=2pt, left=5pt, right=5pt, top=5pt, bottom=5pt]

\small\ttfamily
You are an expert relevance assessor. Given a search query and two passages, state which passage is more relevant to the query. Answer with a single letter: `A' or `B'.

\

Search Query: \{query\} \\
Passage A: \{passage1\} \\
Passage B: \{passage2\} \\
Which passage is more relevant? (A/B)

\end{tcolorbox}

\caption{The prompt used in our pairwise preference experiments.}
\label{fig:preference-prompt}
\end{figure}

\begin{figure}[t]
\centering
\begin{tikzpicture}[
  label/.style={anchor=east, minimum height=0.7cm, font=\ttfamily\footnotesize},
  lbox/.style={rectangle, draw=black, minimum width=0.7cm, minimum height=0.7cm, font=\ttfamily\scriptsize, align=left},
  rbox/.style={rectangle, draw=black, minimum width=3.8cm, minimum height=0.7cm, font=\ttfamily\scriptsize, align=left},
  arrow/.style={-Stealth, thick, shorten >=2pt, shorten <=2pt},
  node distance=0.2cm
]

\node[label] (label_1) {\textbf{Passage A:}};
\node[lbox, right=of label_1] (bef_1) {\textbf{P1}};
\node[rbox, right=1.9cm of bef_1] (inj_1) {\textbf{\textcolor{red}{Published on 1980/01/01.} P1}};

\node[label, below=of label_1] (label_2) {\textbf{Passage B:}};
\node[lbox, below=of bef_1] (bef_2) {\textbf{P2}};
\node[rbox, below=of inj_1] (inj_2) {\textbf{\textcolor{red}{Published on 2025/01/01.} P2}};

\draw[arrow] (bef_1.east) -- node[midway, above=1pt,
                font=\footnotesize\bfseries]{Date injection} (inj_1.west);
\draw[arrow] (bef_2.east) -- (inj_2.west);

\end{tikzpicture}

\caption{Date injection strategy: An old date (1980/01/01) is injected before the preferred passage, and a fresh date is assigned to the other one. In this example, Passage A is preferred before date injection.}
\label{fig:date-injection-preference}
\end{figure}

\section{Evaluation Metrics and Experiment Results}
To our knowledge, no prior work defines metrics expressly for quantifying recency bias in reranking. We therefore introduce a suite of measures that capture the impact of date injection at both the SERP and SERP-Segment levels. For every metric that is a mean over a topic set, we apply a two-sided one-sample \textit{t-test} on the differences to test whether their mean equals zero (equivalently, a two-sided paired \textit{t-test}) and treat results as statistically significant when $p < 0.05$. Unless noted otherwise, we report each metric's mean across all topics and supplement this with per-topic Kendall's~tau to gauge model robustness.

\subsection{Absolute Average and Largest Rank Shift}
To quantify how much individual ranks change after date injection for each query, we introduce two SERP-level metrics: \textbf{Absolute Average Rank Shift (AARS)} and \textbf{Absolute Largest Rank Shift (ALRS)}. By taking the absolute value of shifts regardless of direction, we can see the average impact on the entire SERP. Let $r_i$ and $r_i^{\text{inj}}$ denote the rank of document~$i$ before and after date injection, respectively. The rank shift for document~$i$ is defined as:

\begin{equation}
\Delta r_i = r_i^{\text{inj}} - r_i.
\label{eq:delta_rank}
\end{equation}

Then, the \textbf{Absolute Average Rank Shift (AARS)} is the mean absolute displacement across the SERP:

\begin{equation}
\text{AARS} = \frac{1}{N} \sum_{i=1}^{N} \left| \Delta r_i \right|,
\label{eq:aars}
\end{equation}
where $N$ is the total number of documents.

The \textbf{Absolute Largest Rank Shift (ALRS)} reports the greatest absolute displacement observed across all documents:

\begin{equation}
\text{ALRS} = \max_{i \in \{1, \dots, N\}} \left| \Delta r_i \right|.
\label{eq:alrs}
\end{equation}

For collection-level summaries, we compute the \emph{mean} AARS (\textbf{\emph{m}AARS}) and the \emph{maximum} ALRS across all topics ($\mathbf{ALRS}_{\text{all}}$) for each test collection. Let $T$ be the total number of topics (queries) and let $\text{AARS}_t$ and $\text{ALRS}_t$ denote the AARS and ALRS values computed for topic $t$:

\begin{equation}
\text{\emph{m}AARS} = \frac{1}{T}\sum_{t=1}^{T} \text{AARS}_t,
\label{eq:maars}
\end{equation}

\begin{equation}
\text{ALRS}_{all} = \max_{t \in \{1, \dots, T\}} \text{ALRS}_t ,
\label{eq:alrs_all}
\end{equation}

A high \emph{m}AARS signals overall volatility, whereas a high $\text{ALRS}_{all}$ exposes extreme per-passage shifts. In both metrics, lower values indicate greater resistance to recency bias. Table~\ref{tab:rank_shift_metrics} summarises the results. GPT-4o achieves the lowest \emph{m}AARS on both DL21 (1.8204) and DL22 (2.0047), indicating the strongest overall robustness to temporal cues. In contrast, LLaMA3-8B is the most volatile (5.0008 and 5.2782, respectively). Notably, GPT-4o outperforms GPT-4 in \emph{m}AARS, despite GPT-4 being the most expensive model evaluated. 

Turning to $\text{ALRS}_{all}$, the models that excel in \emph{m}AARS also tend to exhibit greater robustness under extreme shifts. However, even the best case (Qwen2.5-7B on DL21) still suffers the single largest shift of 61 positions, confirming that none of the tested LLMs are immune to date-injection perturbations.

All results of \emph{m}AARS shown in Table~\ref{tab:rank_shift_metrics} are statistically significant $(p < 0.05)$, showing that date injection leads to systematic recency bias for every LLM.

\begin{table}[t]
  \centering
    \caption{Mean Absolute Average Rank Shift (\emph{m}AARS) and Absolute Largest Rank Shift Over All Topics ($\text{ALRS}_{all}$) on DL21 and DL22. A lower value suggests the model is more robust. Red values were negative before taking the absolute value. All results of \emph{m}AARS are statistically significant at $p < 0.05$. The $p$-value is obtained from a \textit{t-test}.}
    \label{tab:rank_shift_metrics}
    \begin{tabular}{lcccc}
        \toprule
        \multirow{2}{*}{\textbf{Model}} & \multicolumn{2}{c}{\textbf{DL21}} & \multicolumn{2}{c}{\textbf{DL22}} \\
        \cmidrule(lr){2-3} \cmidrule(lr){4-5}
        & \textbf{\emph{m}AARS} & \textbf{$\text{ALRS}_{all}$} & \textbf{\emph{m}AARS} & \textbf{$\text{ALRS}_{all}$} \\
        \midrule
        GPT-3.5-turbo            & 3.5811 & \negcell{95} & 3.7537 & 85 \\
        GPT-4o                   & 1.8204 & 70 & 2.0047 & \negcell{79} \\
        GPT-4                    & 2.0660 & 69 & 2.3126 & \negcell{86} \\
        LLaMA3-8B                & 5.0008 & 93 & 5.2782 & \negcell{89} \\
        LLaMA3-70B               & 2.6125 & 82 & 2.4234 & \negcell{83} \\
        Qwen2.5-7B               & 3.5385 & 61 & 3.6871 & \negcell{81} \\
        Qwen2.5-72B              & 1.9166 & 77 & 2.2729 & 87 \\
        \bottomrule
    \end{tabular}
\end{table}

\subsection{Average Year Shift in Top-K}
To examine the effect on top-ranked results that matter most to users, we introduce a SERP-segment-level metric, \textbf{Average Year Shift in Top-K} (denoted $YS^{(K)}$). Let $y_i^{\text{before}}$ and $y_i^{\text{after}}$ be the publication years of the passage at Rank~$i$ before and after date injection, respectively, with $y_i^{\text{before}}$ being the injected publication year. For a given cutoff $K$, we define

\begin{align}
YS^{(K)} &= \frac{1}{K} \sum_{i=1}^{K} \left( y_i^{\text{after}} - y_i^{\text{before}} \right). 
\label{eq:delta_ys_k}
\end{align}

The collection-level mean is

\begin{align}
mYS^{(K)} &= \frac{1}{T}\sum_{t=1}^{T}  YS_t^{(K)}. 
\label{eq:m_delta_ys_k}
\end{align}

\begin{table}[ht]
  \centering
  \caption{Mean year shift in top-K ranked passages before and after date injection, on DL21 and DL22. A lower value suggests the model is more robust. All results are statistically significant at $p < 0.05$. The $p$-value is obtained from a \textit{t-test}.}
  \label{tab:topk_mean_year_shift_combined}
  \begin{tabular}{llcccc}
    \toprule
    \multirow{2}{*}{\textbf{Model}} & & \multicolumn{4}{c}{\textbf{$mYS^{(K)}$}}\\
    \cmidrule{3-6}
    & & \textbf{K = 10} & \textbf{20} & \textbf{30} & \textbf{50} \\
    \midrule
    \multirow{2}{*}{GPT-3.5-turbo}
                & DL21 & 3.238 & 2.058 & 1.577 & 0.896 \\
                & DL22 & 2.968 & 1.793 & 1.430 & 0.860 \\
    \midrule
    \multirow{2}{*}{GPT-4o}
                & DL21 & 1.300 & 0.742 & 0.721 & 0.445 \\
                & DL22 & 1.400 & 1.100 & 0.881 & 0.536 \\
    \midrule
    \multirow{2}{*}{GPT-4}
                & DL21 & 1.323 & 0.863 & 0.752 & 0.383 \\
                & DL22 & 1.863 & 1.253 & 1.057 & 0.616 \\
    \midrule
    \multirow{2}{*}{LLaMA3-8B}
                & DL21 & 3.908 & 2.367 & 1.808 & 1.042 \\
                & DL22 & 4.780 & 2.774 & 1.929 & 1.253 \\
    \midrule
    \multirow{2}{*}{LLaMA3-70B}
                & DL21 & 2.800 & 1.549 & 1.042 & 0.806 \\
                & DL22 & 2.176 & 1.518 & 1.143 & 0.695 \\
    \midrule
    \multirow{2}{*}{Qwen2.5-7B}
                & DL21 & 2.049 & 1.511 & 1.213 & 0.595 \\
                & DL22 & 2.792 & 1.683 & 1.189 & 0.843 \\
    \midrule
    \multirow{2}{*}{Qwen2.5-72B}
                & DL21 & 0.819 & 0.608 & 0.488 & 0.323 \\
                & DL22 & 1.462 & 1.031 & 0.749 & 0.397 \\
    \bottomrule
  \end{tabular}
\end{table}

Unlike \emph{m}AARS, we \emph{do not} take absolute values here, because the metric captures the average year shift within the top-K segment and our experimental design guarantees the segment has the lowest possible average year before date injection. Any values above zero indicate that the model favours newer passages, pulling fresher candidates into the top-K. Table~\ref{tab:topk_mean_year_shift_combined} reports results for $K\!\in\!\{10,20,30,50\}$, in which all values are statistically significant at $p < 0.05$, confirming our date-injection strategy consistently drives every LLM to promote passages with newer timestamps.

GPT-4o and Qwen2.5-72B again prove most robust, while LLaMA3-8B is the most recency-sensitive, making the top-10 on average 4.780 years newer. All models show smaller shifts as $K$ grows. For LLaMA3-8B, on average, each of the top-10 passages becomes 3.908 years fresher on DL21 and 4.780 years fresher on DL22. In comparison, the strongest models limit the shift to 0.819 years (Qwen2.5-72B on DL21) and 1.400 years (GPT-4o on DL22). When the cutoff widens to the top-50, the effect diminishes across the board, yet LLaMA3-8B still renders the list about a year newer (1.042 and 1.253 years). We attribute this attenuation at larger K to the dilution of extreme rank changes once the tail of the SERP is included.

\subsection{Average Year Shift by Groups}

To further analyse the effect, we introduce another SERP-segment-level metric, \textbf{Average Year Shift by Groups} ($YSG^{(g)}$), which quantifies temporal drift across different portions of the ranking. Each ranked list is divided into deciles (groups of ten). As with $YS^{(K)}$, we \emph{do not} take absolute values. In this setting, the average year for the middle segments, every decile except the first and last, can move in either direction. For the $g$-th ($g=0, \ldots, 9$) group, covering positions $[10g+1, 10g+10]$, we define the average year shift as:


\begin{align}
YSG^{(g)} &= \frac{1}{|G_g|} \sum_{i \in G_g} \left( y_i^{\text{after}} - y_i^{\text{before}} \right), \label{eq:group_mean_year_shift}
\end{align}

where $G_g = \{ i \mid 10g + 1 \leq \text{rank}_i < 10g + 11 \}$ denotes the set of documents in group $g$ based on their respective ranks. Grouping is applied independently before and after reranking.

Again, we report the mean value:

\begin{align}
mYSG^{(g)} &= \frac{1}{T}\sum_{t=1}^{T} YSG_t^{(g)}. 
\label{eq:m_group_mean_year_shift}
\end{align}

\begin{table*}[ht]
  \centering
  \caption{Mean publication year shift by rank groups before and after date injection, on DL21 and DL22. Positive values are shown in bold. Entries marked with \sig are statistically significant at $p < 0.05$. The $p$-value is obtained from a \textit{t-test}.}
  \label{tab:group_mean_year_shift_combined}
  \resizebox{\linewidth}{!}{
  \begin{tabular}{llcccccccccc}
    \toprule
    \multirow{2}{*}{\textbf{Model}} &  & \multicolumn{10}{c}{\textbf{$mYSG^{(g)}$}}\\
    \cmidrule{3-12}
    & & \textbf{1--10} & \textbf{11--20} & \textbf{21--30} & \textbf{31--40} & \textbf{41--50}
                   & \textbf{51--60} & \textbf{61--70} & \textbf{71--80} & \textbf{81--90} & \textbf{91--100} \\
    \midrule
    \multirow{2}{*}{GPT-3.5-turbo}
                 & DL21 & \textbf{+3.238\sig} & \textbf{+0.879\sig} & \textbf{+0.613} & \textbf{+0.274} & -0.525 & -0.647 & -0.238 & -1.200\sig & -1.087\sig & -1.308\sig \\
                 & DL22 & \textbf{+2.968\sig} & \textbf{+0.618\sig} & \textbf{+0.703\sig} & -0.089 & \textbf{+0.099} & \textbf{+0.301} & -0.268 & -1.137\sig & -1.570\sig & -1.625\sig \\
    \midrule
    \multirow{2}{*}{GPT-4o}
                 & DL21 & \textbf{+1.300\sig} & \textbf{+0.183} & \textbf{+0.679\sig} & \textbf{+0.089} & -0.028 & -0.072 & -0.515 & -0.458 & -0.508\sig & -0.670\sig \\
                 & DL22 & \textbf{+1.400\sig} & \textbf{+0.800\sig} & \textbf{+0.442} & \textbf{+0.118} & -0.082 & -0.011 & -0.749\sig & -0.578\sig & -0.674\sig & -0.668\sig \\
    \midrule
    \multirow{2}{*}{GPT-4}
                 & DL21 & \textbf{+1.323\sig} & \textbf{+0.404} & \textbf{+0.530\sig} & \textbf{+0.128} & -0.470 & -0.057 & -0.621\sig & -0.040 & -0.496 & -0.702\sig \\
                 & DL22 & \textbf{+1.863\sig} & \textbf{+0.643\sig} & \textbf{+0.663\sig} & -0.011 & -0.082 & -0.258 & -0.357 & -0.724\sig & -0.792\sig & -0.947\sig \\
    \midrule
    \multirow{2}{*}{LLaMA3-8B}
                 & DL21 & \textbf{+3.908\sig} & \textbf{+0.826\sig} & \textbf{+0.691} & \textbf{+0.783} & -0.449 & -0.740 & -0.732\sig & -1.226\sig & -1.192\sig & -1.868\sig \\
                 & DL22 & \textbf{+4.780\sig} & \textbf{+0.767\sig} & \textbf{+0.239} & \textbf{+0.457} & \textbf{+0.021} & -0.528 & -0.388 & -1.676\sig & -1.704\sig & -1.968\sig \\
    \midrule
    \multirow{2}{*}{LLaMA3-70B}
                 & DL21 & \textbf{+2.800\sig} & \textbf{+0.298} & \textbf{+0.026} & \textbf{+0.951\sig} & -0.045 & -0.430 & -1.006\sig & -0.974\sig & -0.743\sig & -0.877\sig \\
                 & DL22 & \textbf{+2.176\sig} & \textbf{+0.861\sig} & \textbf{+0.392} & \textbf{+0.121} & -0.074 & -0.367 & -0.670\sig & -0.653\sig & -0.764\sig & -1.022\sig \\
    \midrule
    \multirow{2}{*}{Qwen2.5-7B}
                 & DL21 & \textbf{+2.049\sig} & \textbf{+0.974\sig} & \textbf{+0.617} & -0.006 & -0.658 & -0.092 & -0.636\sig & -0.625\sig & -0.464\sig & -1.158\sig \\
                 & DL22 & \textbf{+2.792\sig} & \textbf{+0.574\sig} & \textbf{+0.200} & \textbf{+0.650} & -0.003 & -0.774\sig & -0.605 & -0.457 & -1.014\sig & -1.363\sig \\
    \midrule
    \multirow{2}{*}{Qwen2.5-72B}
                 & DL21 & \textbf{+0.819\sig} & \textbf{+0.396} & \textbf{+0.249} & \textbf{+0.043} & \textbf{+0.109} & -0.247 & -0.442 & -0.036 & -0.281 & -0.611\sig \\
                 & DL22 & \textbf{+1.462\sig} & \textbf{+0.600\sig} & \textbf{+0.184} & -0.022 & -0.238 & -0.239 & -0.317 & -0.217 & -0.697\sig & -0.514\sig \\
    \bottomrule
  \end{tabular}
  }
\end{table*}

Table \ref{tab:group_mean_year_shift_combined} shows a clear, consistent trend across all models and both test collections. We note:

\textbf{Top of the list becomes markedly fresher.} Every model shows a statistically significant positive shift in the first decile (ranks 1–10), which is identical to $YS^{(10)}$, and in most cases, the second decile (11–20) as well. Because shifts in the 11–20 band are not guaranteed to be positive, their uniform direction underscores the strength of the recency effect.

\textbf{Middle deciles hover near neutrality.} Shifts in the 21–60 region are generally small and often non-significant. The upper half (21–40) tends to lean slightly positive, whereas the lower half (41–60) is more likely to shift slightly negative, suggesting a pivot point near the SERP's centre.

\textbf{Bottom of the list becomes older.} From the seventh decile onward (61-70, 71–80, 81–90, 91–100), every value is negative; most are significant at $p < 0.05$, with significance growing stronger toward the tail. For most model--collection combinations, the last decile experiences the largest backward shift, up to –1.968 years for LLaMA3-8B on DL22. Even the most robust model, Qwen2.5-72B, records more than a half-year negative shift in the final decile on both datasets.

\textbf{Entire SERP tilts around a pivot near the centre.} Overall, the SERP behaves like a seesaw: passages stamped with recent dates are pulled into the top 40, while older-dated passages slide toward the bottom. The mid-SERP deciles (41–60) act as a pivot, exhibiting the smallest absolute changes.

These findings confirm that date injection systematically elevates newer-dated passages and demotes older ones, with the magnitude of the effect varying by model size and architecture. Considering our experiment design, this trend suggests that older-dated passages are more difficult to be promoted within the reranking window.

\subsection{Kendall's tau}
\begin{figure}[ht]
  \centering
  \includegraphics[width=\linewidth]{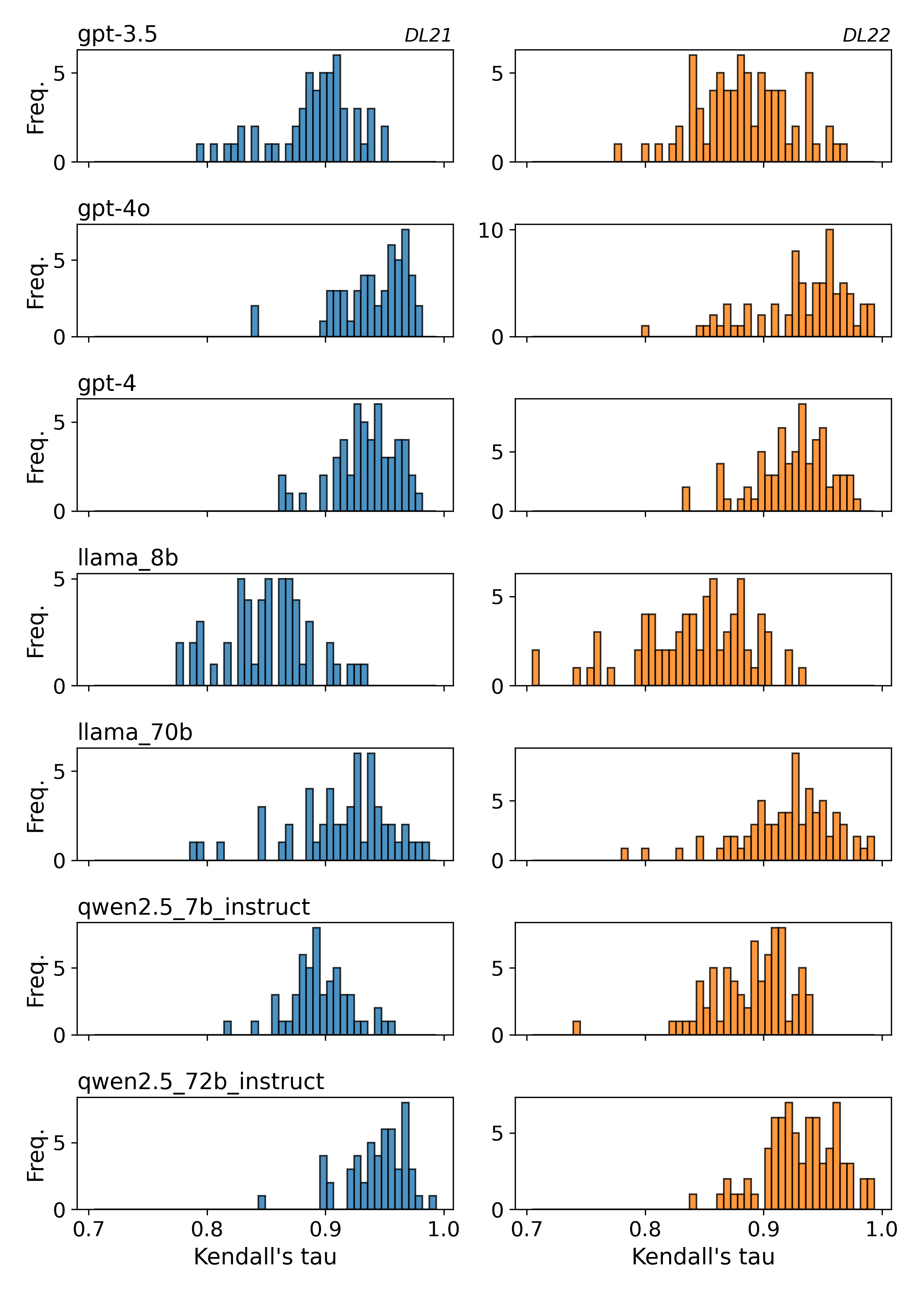}
  \caption{Kendall's tau distributions for each model on DL21 (left) and DL22 (right).}
  \label{fig:kendall}
\end{figure}

To assess how strongly date injection reshapes the overall ranking, we compute Kendall's~tau between the two reranked SERPs for each query by comparing the document IDs. Figure~\ref{fig:kendall} shows the resulting distributions (DL21 on the left and DL22 on the right), from which we can see a clear trend: larger, more capable LLMs produce rankings that remain more stable after date injection (higher Kendall's~tau), whereas smaller models exhibit greater sensitivity to recency cues (lower Kendall's~tau). The effect is particularly evident within each provider's line-up. For example, LLaMA3-70B surpasses LLaMA3-8B, and Qwen2.5-72B outperforms Qwen2.5-7B. Likewise, both GPT-4 and GPT-4o achieve higher Kendall's~tau than GPT-3.5-turbo, with GPT-4o again edging out GPT-4. These results align with our earlier metrics, confirming that larger models are generally more resistant to recency bias.

\subsection{Reversal Rate}
For pairwise preference experiments, we compute the \textbf{reversal rate (RR)} for each relevance level $r\in\{0,1,2,3\}$ and topic $t$:
\[
  RR_{r}^{(t)}
  \;=\;
  \frac{\#\,\text{reversed pairs}}
       {\#\,\text{evaluated pairs}}
  \;\in[0,1].
\]

Only pairs with valid preference results both before and after date injection are included in the RR calculation. We then report the \emph{mean} and \emph{maximum} RR across all topics where at least one corresponding valid pair exists. Let $T_r$ denote the number of such topics at relevance level $r$.
\begin{align*}
  \overline{RR}_{r} &=
  \frac{1}{T_r}\sum_{t=1}^{T_r}RR_{r}^{(t)}, &
  RR^{\max}_{r} &= 
  \max_{1 \leq t \leq T_r} RR_r^{(t)},
\end{align*}

All results of $\overline{RR}_{r}$ in Table~\ref{tab:rr-by-llm} are statistically significant, reinforcing the earlier trend that large LLMs are generally more resistant to date-injection perturbations, though the effect varies by relevance level. LLaMA3-8B is the most vulnerable overall, with an $\overline{RR}_{\text{All}}$ of 25.18\%. By contrast, LLaMA3-70B proves more sensitive when the passages are highly relevant (level 2), posting the highest $\overline{RR}_{2}$ (29.63\%) and $RR^{\max}_{2}$ (81.02\%). For perfectly relevant (level 3) pairs, since many topics do not have many pairs and some topics only have one pair, the results of $RR^{\max}_{3}$ are not very informative. For LLaMA3-8B, LLaMA3-70B, and Qwen2.5-7B, the $RR^{\max}_{3}$ is 100\%. However, these results are measured in the case where only one pair exists. Also, it should be noted that only 40 out of 53 topics have at least one level 3 pair. For these reasons, our subsequent discussion of Max RR focuses on relevance levels 0, 1, and 2.

Overall, the Qwen2.5 models are noticeably more robust than the LLaMA3 models. Qwen2.5-72B is the least affected, with an $\overline{RR}_{\text{All}}$ of just 8.93\%. Remarkably, even the smaller model, Qwen2.5-7B, at only 1/10 the parameters of LLaMA3-70B, outperforms LLaMA3-70B across every relevance level.

\begin{table}[ht]
  \centering
  \caption{Reversal rate (RR) after date injection. Higher indicates stronger temporal bias. Relevance levels are coded as 0 = irrelevant, 1 = related, 2 = highly relevant, and 3 = perfectly relevant. For ``All'', we pool valid pairs across relevance levels within each topic, then report the mean and maximum of the resulting topic-level RRs. All results of Mean RR ($\overline{RR}_{r}$) are statistically significant at $p < 0.05$. The $p$-value is obtained from a \textit{t-test}.}
  \label{tab:rr-by-llm}
  \setlength{\tabcolsep}{3pt}
  \begin{tabular}{llcc}
    \toprule
    \textbf{Model} & \textbf{Relevance} & \textbf{Mean RR ($\overline{RR}_{r}$)} & \textbf{Max RR ($RR^{\max}_{r}$)} \\
    \midrule
    \multirow{5}{*}{LLaMA3-8B}     & 0 & 0.2191 & 0.4975 \\
                                   & 1 & 0.2785 & 0.5915 \\
                                   & 2 & 0.2366 & 0.7431 \\
                                   & 3 & 0.2433 & 1.0000 \\
                                   & All & 0.2518 & 0.6045 \\
    \midrule
    \multirow{5}{*}{LLaMA3-70B}    & 0 & 0.1131 & 0.2941 \\
                                   & 1 & 0.2454 & 0.5286 \\
                                   & 2 & 0.2963 & 0.8102 \\
                                   & 3 & 0.2804 & 1.0000 \\
                                   & All & 0.2102 & 0.6302 \\
    \midrule
    \multirow{5}{*}{Qwen2.5-7B}    & 0 & 0.0986 & 0.2098 \\
                                   & 1 & 0.1341 & 0.2571 \\
                                   & 2 & 0.1305 & 0.4286 \\
                                   & 3 & 0.1437 & 1.0000 \\
                                   & All & 0.1210 & 0.2892 \\
    \midrule
    \multirow{5}{*}{Qwen2.5-72B}   & 0 & 0.0575 & 0.1434 \\
                                   & 1 & 0.0847 & 0.1611 \\
                                   & 2 & 0.1128 & 0.2955 \\
                                   & 3 & 0.1231 & 0.4000 \\
                                   & All & 0.0893 & 0.2104 \\
    \bottomrule
  \end{tabular}
\end{table}

\section{Discussion}
Our listwise reranking experiments provide clear empirical evidence that LLMs display a measurable \textbf{recency bias} when employed as listwise rerankers. Injecting a single artificial publication date, without altering any semantic content, is sufficient to induce sizeable shifts in the ranked lists across two TREC passage retrieval test collections. Pairwise preference tests align with this finding: a simple date tag can flip an LLM's judgement of which passage is ``more relevant''.

\subsection{Magnitude and Trend of Recency Bias}

The results of $mYS^{(K)}$ (a SERP-segment-level metric) in Table~\ref{tab:topk_mean_year_shift_combined} indicate that all seven models favour newer content. Even the most robust model (Qwen2.5-72B) pushes the upper half of the SERP forward by 0.323 years on DL21 and 0.397 years on DL22, while the least robust (LLaMA3-8B) advances the same segment by over one year, 1.042 years on DL21 and 1.253 years on DL22. The SERP-level metrics tell a consistent story: list-wide volatility (\emph{m}AARS) is non-negligible, and extreme rank shifts (\( \text{ALRS}_{\text{all}} \)) appear for every model.

Another SERP-segment-level metric in Table~\ref{tab:group_mean_year_shift_combined} reinforces this trend. The top four deciles almost always become younger, while those demoted to the tail skew older. This trend confirms that the injected timestamps are interpreted by LLMs as a strong, largely monotonic relevance signal that overrides other term-based and semantics-based evidence.

Pairwise experiment results echo this trend. On LLaMA3-8B, more than 25\% of baseline preferences flip after date injection; even the most robust model (Qwen2.5-72B) shows per-topic reversal rates as high as 29.55\% for highly relevant (level 2) pairs. Preference reversals occur at every relevance level, with all four models showing their lowest Mean RR on irrelevant passages (level 0). Mean RR peaks at level 3 for both Qwen2.5 models, at level 1 for LLaMA3-8B, and at level 2 for LLaMA3-70B.

\subsection{Model Capacity and Robustness}
Overall, larger models tend to be more resistant to recency bias. From Table~\ref{tab:rank_shift_metrics}, we can see large models (GPT-4o, GPT-4, LLaMA3-70B, Qwen2.5-72B) exhibit markedly lower \emph{m}AARS than smaller counterparts (GPT-3.5-turbo, LLaMA3-8B, Qwen2.5-7B) from the same provider. GPT-4o, for example, averages 1.8204 (DL21) and 2.0047 (DL22), whereas GPT-3.5-turbo averages 3.5811 and 3.7537, respectively. However, no model is immune. The results of $\text{ALRS}_{\text{all}}$ exhibit extreme rank shifts that occur for all LLMs. Even the best case (Qwen2.5-7B on DL21) still contains a 61-rank shift. Kendall's~tau distributions reinforce the same message. The pairwise tests also show lower overall Mean RR for the larger model in each family, although LLaMA3-70B has higher Mean RR than LLaMA3-8B at relevance levels 2 and 3. Recency bias is therefore a pervasive weakness across today's LLM-based rerankers.

\subsection{Evaluation Metrics: SERP vs.~SERP-Segment Level Perspective}
Our metric families complement each other. \textbf{SERP-level rank-shift metrics} (\emph{m}AARS/$\text{ALRS}_{all}$) capture overall stability, while \textbf{SERP-Segment-level year-shift metrics} ($mYS^{(K)}$, $mYSG^{(g)}$) reveal the direction and location of temporal shifts. A model can post a moderate \emph{m}AARS yet still show a large $mYS^{(K)}$ if changes are concentrated at the very top of the list. Indeed, LLaMA3-70B outperforms Qwen2.5-7B on \emph{m}AARS but matches or even underperforms it on $mYS^{(K)}$ for K = 10, 20, and 50 on DL21. Table~\ref{tab:group_mean_year_shift_combined} further underscores a clear ``seesaw'' pattern: the influence of temporal signals grows stronger toward both ends of the SERP.

\subsection{Limitations and Future Work}

Although our controlled experiments clearly expose a recency preference, they remain limited in scope. We focus on two TREC collections; however, incorporating additional and more diverse datasets would provide a broader picture. Pairwise tests are restricted to four open-source models and a single test collection (DL21) due to cost constraints. Extending to more LLMs, including proprietary models, and other test collections, is important future work. 

Methodologically, our listwise setup fixes the sliding-window size. The trend in Table~\ref{tab:group_mean_year_shift_combined} suggests that varying the window size or exploring alternative chunking strategies could influence how strongly timestamps sway the model and therefore deserves systematic study. Likewise, richer manipulations (e.g. ``Breaking news'', ``Updated today'' tags) and mixed-relevance passage pairs could deepen our understanding, though at a higher computational cost. The size of the time gap itself may modulate the strength of recency bias as well. We leave a thorough study of this factor to future work.

Furthermore, analysing the underlying causes of recency bias is an important direction for future work. This bias may be learned from pre-training data, though we cannot rule out that it is an emergent LLM behaviour.

\section{Conclusion}
This study set out to answer a straightforward but long-overlooked question: \emph{Do LLMs systematically prefer newer content when acting as search rerankers?} Through listwise experiments on the DL21 and DL22 passage retrieval collections, we uncover a pronounced \textbf{recency bias}. All seven models, spanning proprietary (GPT-3.5-turbo, GPT-4, GPT-4o) and open-source (LLaMA3-8B/70B, Qwen2.5-7B/72B), systematically promote passages with newer timestamps, pushing the average publication year of the top-10 results forward by up to 4.780 years and moving individual items by as many as 95 ranks. Pairwise tests on DL21 reinforce the finding: a simple date tag can reverse up to 25\% of preferences between equally relevant passages. While larger models attenuate the effect, none eradicate it, confirming that recency bias is systemic, rather than merely a small-model quirk.

By isolating timestamps as the sole perturbation, we provide the quantitative evidence that recency functions as an implicit relevance signal for LLM-based rerankers. This exposes a concrete risk: if temporal cues go unchecked, LLM-based rerankers may undervalue authoritative yet older material, which is problematic in domains where historical evidence should weigh as heavily as recent information. Recency bias is almost certainly just one of many latent biases still hidden from view. We therefore urge the IR community to broaden the bias map beyond recency and to develop mitigation strategies that keep future LLM-centric retrieval systems robust against such distortions.

\begin{acks}
We thank the members of The Sakai Laboratory for helpful discussions. We are grateful to the anonymous reviewers and the SIGIR-AP 2025 organizers for their constructive feedback. We also thank our families and friends for their unwavering support.
\end{acks}

\bibliographystyle{ACM-Reference-Format}
\bibliography{recency_bias_llm}

\appendix

\end{document}